\documentclass[aps,prb,twocolumn,superscriptaddress]{revtex4-2}

\usepackage{amsmath,amssymb}
\usepackage{graphicx}
\usepackage{color}
\usepackage{rotating}
\usepackage{bm}
\usepackage{setspace}
\usepackage{hyperref}
\usepackage{multirow}

\hypersetup{
colorlinks=true,
urlcolor= blue,
citecolor=blue,
linkcolor= blue,
bookmarks=true,
bookmarksopen=false,
}

\begin{document}


\title{Probing the interlayer coupling in 2$H$-NbS$_2$ via \\ soft-x-ray angle-resolved photoemission spectroscopy}


\author{D. Huang}
\email{D.Huang@fkf.mpg.de}
\affiliation{Max Planck Institute for Solid State Research, 70569 Stuttgart, Germany}
\author{H. Nakamura}
\email{hnakamur@uark.edu}
\affiliation{Department of Physics, University of Arkansas, Fayetteville, Arkansas 72701, USA}
\author{K. K\"{u}ster}
\affiliation{Max Planck Institute for Solid State Research, 70569 Stuttgart, Germany}
\author{U. Wedig}
\affiliation{Max Planck Institute for Solid State Research, 70569 Stuttgart, Germany}
\author{N. B. M. Schr\"{o}ter}
\altaffiliation{Present address: Max Planck Institute for Microstructure Physics, 06120 Halle, Germany}
\affiliation{Swiss Light Source, Paul Scherrer Institute, CH-5232 Villigen PSI, Switzerland}
\author{V. N. Strocov}
\affiliation{Swiss Light Source, Paul Scherrer Institute, CH-5232 Villigen PSI, Switzerland}
\author{U. Starke}
\affiliation{Max Planck Institute for Solid State Research, 70569 Stuttgart, Germany}
\author{H. Takagi}
\affiliation{Max Planck Institute for Solid State Research, 70569 Stuttgart, Germany}
\affiliation{Institute for Functional Matter and Quantum Technologies, University of Stuttgart, 70569 Stuttgart, Germany}
\affiliation{Department of Physics, University of Tokyo, 113-0033 Tokyo, Japan}

\date{\today}

\begin{abstract}
In the large family of two-dimensional (2D) layered materials including graphene, its honeycomb analogs, and transition-metal dichalcogenides, the interlayer coupling plays a rather intriguing role. On the one hand, the weak van der Waals interaction that holds the layers together endows these compounds with quasi-2D properties, which might imply small interlayer effects on the electronically active bands. On the other hand, the oft-witnessed differences in electronic, optical, and magnetic behaviors of monolayers, bilayers, and multilayers of the same compound must have as their microscopic origin the detailed interlayer hopping parameters. Given the few experimental reports that have attempted to explicitly extract these parameters, we employ soft-x-ray angle-resolved photoemission spectroscopy (SX-ARPES) to probe the interlayer coupling in superconducting 2$H$-NbS$_2$. We visualize the S 3$p_z$ bands that disperse with respect to the out-of-plane momentum and introduce a simple tight-binding model to extract the interlayer hopping parameters. From first-principles calculations, we clarify how atomic distances and the proper accounting for screening via hybrid functionals influence these bands. The knowledge of interlayer hopping parameters is particularly pertinent in NbS$_2$, where recent experiments have uncovered fingerprints of finite-momentum superconductivity in the bulk material and heterostructures.
\end{abstract}


\maketitle

\section{Introduction}

The broad class of van der Waals (vdW) materials consisting of two-dimensional (2D) atomic layers weakly bonded in the third dimension continues to fuel intense research activity. The ability to isolate monolayers with properties that are sharply distinct from the bulk is particularly appealing. For example, when a single sheet of graphene is exfoliated from graphite, a 2D Dirac semimetal with zero band gap is obtained \cite{Novoselov_Science_2004}. When semiconducting MoS$_2$ is thinned down to a monolayer, the band gap changes from indirect to direct \cite{Mak_PRL_2010}. A bilayer of CrI$_3$ is antiferromagnetic, but a monolayer (or an odd number of layers) is ferromagnetic \cite{Huang_Nature_2017}. Bulk $T_d$-WTe$_2$ is a type-II Weyl semimetal, whereas monolayer $T’$-WTe$_2$ is a 2D topological insulator \cite{Fei_NatPhys_2017, Tang_NatPhys_2017}.

The microscopic origin of these layer-dependent properties is the intricate interlayer coupling in vdW materials. Fundamentally, the problem reduces to deriving universal and transferable forms of interlayer hopping, for example, between the chalcogen $p_z$ orbitals of transition-metal dichalcogenides (TMDCs) \cite{Cappelluti_PRB_2013, Fang_PRB_2015} or the C $2p_z$-like orbitals of graphite \cite{Fang_PRB_2016}, that depend only on the relative distance and orientations of the orbitals. These interlayer hopping formulas yield predictive power for the stacking of different monolayers to form heterostructures, or with a twist angle between the respective lattices \cite{Fang_PRB_2016, Carr_PRB_2018}. While numerical values for these formulas can be extracted from density functional theory (DFT) calculations, two challenges exist: First, the common approximations to the exchange-correlation functional fail to capture the vdW forces that crucially determine the interlayer distances in these compounds. Second, real materials often suffer from atomic and stacking defects that also affect the average interlayer distance. Hence, experimental means of directly probing the interlayer coupling are highly desirable.

Angle-resolved photoemission spectroscopy (ARPES) is a powerful technique for visualizing band structures. Signatures of interlayer coupling may be detected and quantified through additional splitting or gapping of the band structure compared to that of the single layer, or by tracing the band dispersion with respect to the out-of-plane momentum $k_z$~\cite{Markiewicz_PRB_2005, Ohta_PRL_2007, Nicholson_PRL_2017, Nicholson_PRB_2020}. In the latter case, one must note that in the photoemission process, only the in-plane momentum is conserved, and $k_z$ must be inferred assuming a free-electron-like final state and controlled by tuning the incoming photon energy. In this regard, the use of soft-x-ray (SX) photons generated by a synchrotron confers two advantages over the use of conventional vacuum-ultraviolet photons \cite{Strocov_PRL_2012}: First, the higher photon energy (up to 1--2 keV) results in a larger photoelectron escape depth $\lambda$, and in turn, smaller uncertainty in $k_z$ ($\delta k_z = \lambda^{-1}$, where $\delta k_z$ is the full width at half maximum \cite{Strocov_JESRP_2003}). Second, the higher energy of the final state means that the free-electron approximation works better. One limitation in extracting the interlayer coupling from the $k_z$ dispersion is that the system must be 3D or sufficiently thick; this method would not apply to bilayer or few-layer vdW materials.

\begin{figure}
\includegraphics[scale=1]{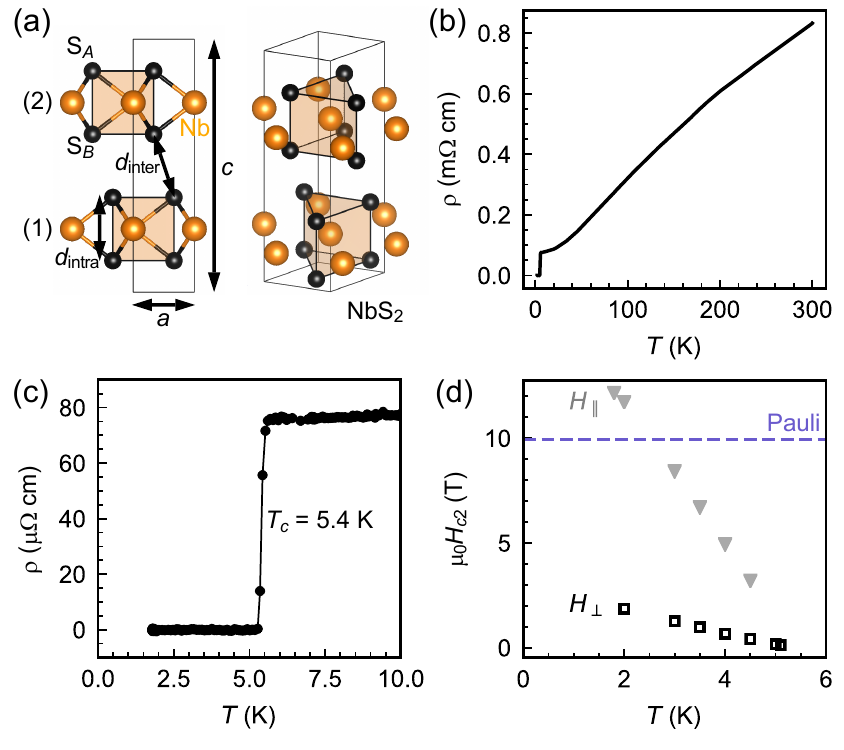}
\caption{(a) Crystal structure and definition of structural parameters. The numbers (1) and (2) and the subscript letters $A$ and $B$ are used in the construction of the TB model [Eq. (\ref{EqH})]. (b) Resistivity vs. temperature from 300 to 1.8 K. (c) Magnification of (b) around the superconducting transition. (d) In-plane ($\parallel$) and out-of-plane ($\perp$) upper critical fields vs. temperature. The former surpasses the theoretical Pauli limit around 2 K.}
\label{Fig1}
\end{figure}

For a model vdW system to explore interlayer coupling, we turn to the superconducting TMDC 2$H$-NbS$_2$ [crystal structure shown in Fig. \ref{Fig1}(a)]. NbS$_2$ is isoelectronic to its more-studied counterpart NbSe$_2$, and even has a similar superconducting transition temperature ($T_c$ $\approx$ 6 K compared to 7 K \cite{Wilson_AP_1975}), yet possesses a few distinct and curious features. The coexisting charge density wave (CDW) phase present in NbSe$_2$ is absent in NbS$_2$ \cite{Naito_JPSJ_1982, Guillamon_PRL_2008, Leroux_PRB_2012, Stan_PRM_2019}, though vestiges may be pinned to defect structures \cite{Leroux_PRB_2018, Wen_PRB_2020}. Theoretical studies point to NbS$_2$ hosting stronger many-body effects than NbSe$_2$, including competing Coulomb and electron-phonon interactions, and lying on the verge of instability to charge or even spin ordering \cite{Nishio_JPSJ_1994, Guller_PRB_2016, Heil_PRL_2017, Heil_PRB_2018, vanLoon_npj_2018, Bianco_NL_2019, Lin_NatComm_2020}. Most surprisingly, when a magnetic field is carefully aligned parallel to the NbS$_2$ layers, the upper critical field shows an upturn above the Pauli limit, reminiscent of the Fulde-Ferrell-Larkin-Ovchinnikov state with finite-momentum Cooper pairs \cite{Cho_NatComm_2021}. However, Ising spin-orbit coupling in TMDC monolayers may provide an alternative means for the system to exceed the Pauli limit, and one deciding factor is the dimensionality of bulk NbS$_2$; i.e., how strong the individual layers are coupled. In this respect, mapping the 3D band structure is foundational.

Here, we report SX-ARPES measurements of NbS$_2$ taken at the ADRESS beamline of the Swiss Light Source \cite{Strocov_JSR_2010, Strocov_JSR_2014}. By tuning the photon energy, we could observe the S 3$p_z$ bands that disperse with respect to $k_z$. We model the $k_z$ dispersion by an effective tight-binding (TB) model, through which the interlayer hopping parameters could be extracted. Extensive DFT calculations elucidate the role of structural parameters and hybrid functionals in reproducing the experimentally observed bands.

\section{Results}

\subsection{Transport}

Figures~\ref{Fig1}(b) and~\ref{Fig1}(c) show resistivity vs. temperature plots of a NbS$_2$ single crystal. Upon cooling, the resistivity exhibits a smooth decrease with no kinks, indicating the absence of a CDW transition in NbS$_2$. At $T_c$ = 5.4 K, a superconducting transition is observed. While the $T_c$ is comparable to literature values, the residual resistivity ratio (RRR) of 11 is slightly lower than that of cleaner single crystals (RRR $\approx$ 20 \cite{Yan_APE_2019} and 70 \cite{Wen_PRB_2020}) and comparable to samples that may be slightly deficient of S \cite{Lian_PhysicaC_2017}. X-ray photoelectron spectroscopy (XPS) measurements revealed an excess of Nb in our samples. These Nb atoms intercalate the vdW gap and create two distinct crystallographic sites for S, evident as two components in the S 2$p$ core levels \cite{SM, Saitoh_JES_2005}. Figure \ref{Fig1}(d) demonstrates the anisotropy between the in-plane and out-of-plane upper critical fields ($H_{c2}$). At 2 K, the ratio $H_{c2, \parallel}/H_{c2, \perp} \approx 6.3$ is similar to reported values of 7.5--8 for NbS$_2$ and larger than reported values of 2.3--3.2 for NbSe$_2$ \cite{Onabe_JPSJ_1978}. $H_{c2, \parallel}$ also exceeds the Pauli limit of 1.84 $T_c$, as previously reported and discussed~\cite{Cho_NatComm_2021}. 

\subsection{SX-ARPES}

\begin{figure*}
\includegraphics[scale=1]{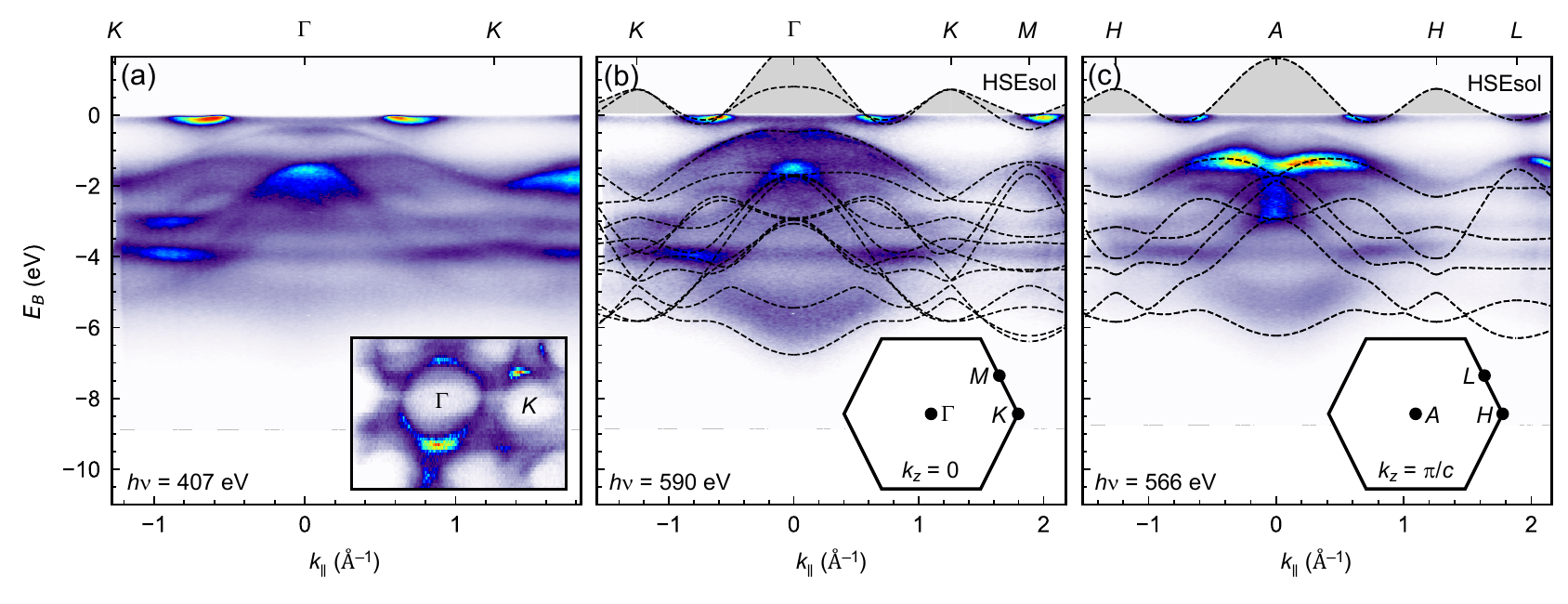}
\caption{In-plane band dispersion of 2$H$-NbS$_2$. (a) and (b) SX-ARPES intensity cuts in the $k_z$ = 0 plane at two photon energies, $h\nu$ = 407 and 590 eV. Inset of (a): Fermi surface in the $k_z$ = 0 plane, visualized by integrating within the binding energy ($E_B$) window [$-$100 meV, 100 meV]; $h\nu$ = 407 eV. A slight artificial distortion is present, due to the cleaved surface exhibiting small flakes and domains with different tilt angles and the challenge of maintaining the beam spot on the same area while tilting the sample platform. (c) SX-ARPES intensity cut in the $k_z$ = $\pi$/$c$ plane; $h\nu$ = 566 eV. In (b) and (c), DFT calculations (HSEsol functional) are overlaid as dashed lines, and the gray-shaded regions mark hole pockets that intersect $E_F$. Insets of (b) and (c): Brillouin zone in the $k_z$ = 0 and $\pi$/$c$ planes.}
\label{Fig2}
\end{figure*}

We begin our discussion of the SX-ARPES results by presenting intensity cuts that span the high symmetry points of the Brillouin zone. Figures \ref{Fig2}(a)--\ref{Fig2}(c) show cuts along the in-plane momentum $k_{\parallel}$ at three different photon energies, $h\nu$ = 407, 590, and 566 eV, corresponding to $k_z$ = 0, 0, and $\pi$/$c$, where $c$ is the length of the unit cell perpendicular to the layers [Fig. \ref{Fig1}(a)]. Circularly polarized photons were used, in order to visualize both orbitals that are odd and even with respect to the $k_x$-$k_z$ plane \cite{SM, Ugeda_NatPhys_2015, Weber_PRB_2018}. The inset of Fig. \ref{Fig2}(a) shows a cut of the Fermi surface at $k_z$ = 0, comprising pockets centered at $\Gamma$ and $K$. As marked by the gray shaded regions in Figs. \ref{Fig2}(b) and \ref{Fig2}(c), these pockets have hole-like nature and extend to the $A$ and $H$ points at $k_z$ = $\pi$/$c$, where according to DFT calculations, the pockets are doubly degenerate. Our results show good agreement with previous ARPES reports of NbS$_2$ probing its 2D band dispersion \cite{Sirica_PRB_2016, Heil_PRB_2018, ElYoubi_PRB_2021}. However, we observe a holelike band at $\Gamma$, just below the Fermi energy ($E_F$), which was absent in the data of Refs. \cite{Sirica_PRB_2016, Heil_PRB_2018}, taken at $h\nu$ = 40 eV. This band was later detected in Ref. \cite{ElYoubi_PRB_2021} at $h\nu$ = 79 eV, but without the full $k_z$ dispersion from 0 to $\pi$/$c$. From DFT calculations, this band comprises S 3$p_z$ orbitals, and its $k_z$ dispersion will play a central role in the ensuing discussion of interlayer coupling.    

\begin{figure}[b]
\includegraphics[scale=1]{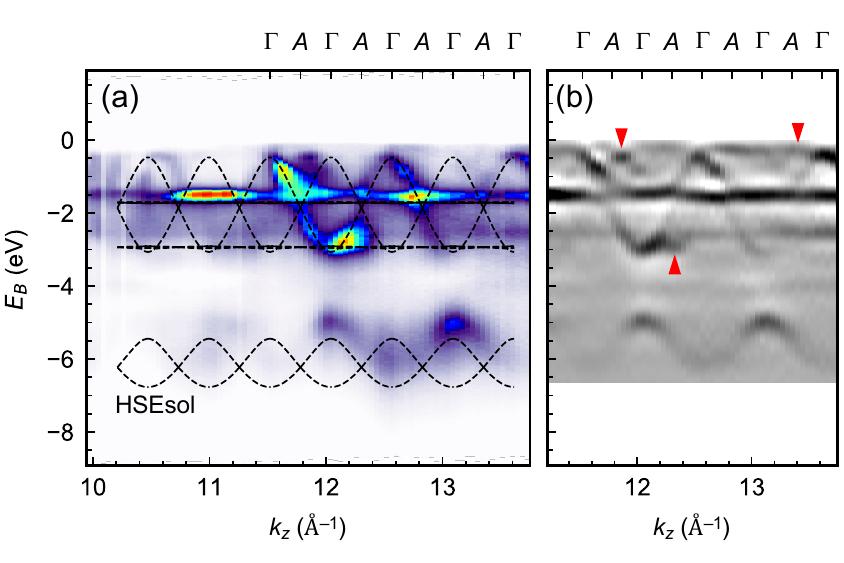}
\caption{Out-of-plane band dispersion of 2$H$-NbS$_2$. (a) SX-ARPES intensity as a function of $k_z$, tuned by the photon energy, along the $\Gamma$--$A$ line ($k_{\parallel}$ = 0). DFT calculations (HSEsol functional) are overlaid as dashed lines. (b) Second-derivative image of (a). The red arrows mark replica bands.}
\label{Fig3}
\end{figure}

We turn our attention to the $k_z$ evolution of the band structure along the $\Gamma$--$A$ line. As seen in Fig. \ref{Fig3}(a) [and more clearly visualized in the second derivative plot; Fig. \ref{Fig3}(b)], there are two pairs of $k_z$-dispersive bands centered around $-1.75$ and $-5$ eV, which should be intimately related to the interlayer coupling, and two sets of $k_z$-nondispersive bands around $-1.5$ and $-2.5$ eV, which are 2D in nature. The $k_z$-dispersive bands exhibit twice the periodicity of the unit cell, an effect which can be attributed to matrix-element effects due to the 2$H$ structure with two NbS$_2$ layers per unit cell \cite{Weber_PRB_2018}. The $k_z$-dispersive bands also show secondary replicas shifted in $k_z$ [marked by red arrows in Fig. \ref{Fig3}(b)], which arise because the final state is not a pure free electron and therefore contains an admixture of other $k_z$ values \cite{Strocov_JESRP_2018}. In spite of these artifacts, the existence of both the $k_z$-dispersive and $k_z$-nondispersive bands is reproduced by DFT calculations [overlaid in Fig. \ref{Fig3}(a)]. The pair of $k_z$-dispersive bands closer to $E_F$ comprises S 3$p_z$ orbitals in an odd combination with respect to the Nb plane, while the other pair of $k_z$-dispersive bands deep below $E_F$ includes both S 3$p_z$ orbitals in an even combination and Nb 4$d_{z^2}$ orbitals. The $k_z$-nondispersive bands closer to $E_F$ comprise four odd combinations of S 3$p_{x/y}$ and Nb 4$d_{xz/yz}$ orbitals with respect to the Nb plane, while the $k_z$-nondispersive bands further below $E_F$ include four even combinations of S 3$p_{x/y}$ and Nb 4$d_{xy/x^2-y^2}$ orbitals. These orbital assignments are consistent with the photon polarization dependence of their ARPES intensities~\cite{SM}.

\begin{figure*}
\includegraphics[scale=1]{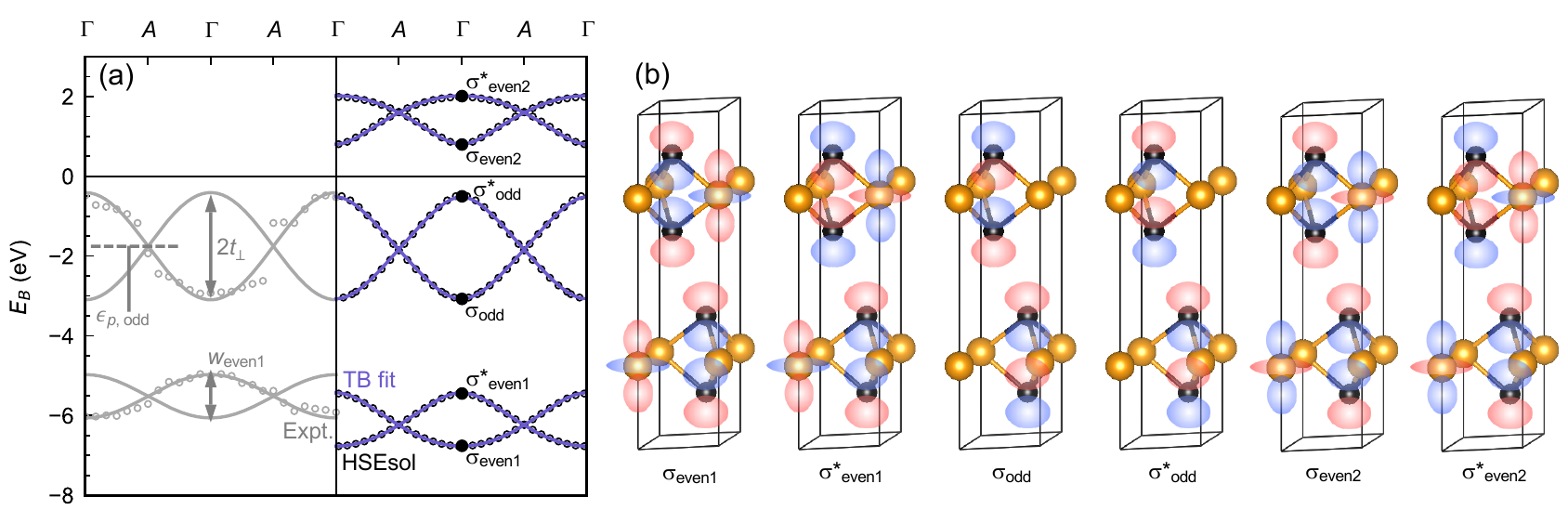}
\caption{Effective model for $k_z$-dispersive bands. (a) Band dispersion along $\Gamma$--$A$ derived from experiment (left), DFT (open circles; HSEsol hybrid functional; right), and a TB fit to DFT (line; Eq. (\ref{EqH}); right). (b) Schematic orbital compositions of the states at $\Gamma$. Red and blue colors denote the phases of the lobes. The odd orbitals S $3p_{z, \mathrm{odd}}$ form one pair of bonding and antibonding states, $\sigma_{\textrm{odd}}$, $\sigma^*_{\textrm{odd}}$, whereas the even orbitals S $3p_{z, \mathrm{even}}$ and Nb $4d_{z^2}$ form two pairs of bonding and antibonding states, $\sigma_{\textrm{even1}}$, $\sigma^*_{\textrm{even1}}$ and $\sigma_{\textrm{even2}}$, $\sigma^*_{\textrm{even2}}$.}
\label{Fig4}
\end{figure*}

\subsection{Effective TB Model for $k_z$-dispersive bands} 

The relative simplicity of the $k_z$-dispersive bands in Fig. \ref{Fig3} motivates the construction of an effective TB model restricted to the $\Gamma$--$A$ line ($k_{\parallel} = 0$). We first take a single NbS$_2$ layer with the 4$d_{z^2}$ orbital of the Nb atom and the 3$p_z$ orbitals of the two S atoms. Since there is mirror symmetry with respect to the Nb plane, the S 3$p_z$ orbitals can be expressed as odd and even combinations that do not mix: $p_{z, \mathrm{odd}} = (p_{z, A} + p_{z, B})/\sqrt{2}$ and $p_{z, \mathrm{even}} = (p_{z, A} - p_{z, B})/\sqrt{2}$, where $p_{z, A}$ and $p_{z, B}$ are orbitals for the S atoms above and below the Nb plane, respectively \cite{Fang_PRB_2015}. Next, we consider the full unit cell with two NbS$_2$ layers [labeled (1) and (2) in Fig. \ref{Fig1}(a)]. Our basis is thus
\begin{equation}
\psi^{\dagger} = (p_{z, \mathrm{odd}}^{(1) \dagger}, p_{z, \mathrm{even}}^{(1) \dagger}, d_{z^2}^{(1) \dagger}, p_{z, \mathrm{odd}}^{(2) \dagger}, p_{z, \mathrm{even}}^{(2) \dagger}, d_{z^2}^{(2) \dagger}).
\end{equation}
Within each layer, the $p_{z, \mathrm{odd}}$, $p_{z, \mathrm{even}}$, and $d_{z^2}$ orbitals have energies $\epsilon_{p, \mathrm{odd}}$, $\epsilon_{p, \mathrm{even}}$, and $\epsilon_{d}$, respectively. We further posit that the only interlayer hopping is between neighboring pairs $p^{(1)}_{z, A}$--$p^{(2)}_{z, B}$ and $p^{(1)}_{z, B}$--$p^{(2)}_{z, A}$, denoted by $t_{\perp}$, and that the only intralayer hybridization allowed by symmetry is between the even orbitals $p_{z, \mathrm{even}}$ and $d_{z^2}$ orbitals, denoted by $t_{\parallel}$. We arrive at the following 6 $\times$ 6 Hamiltonian matrix describing the $k_z$ dispersion along $\Gamma$--$A$:
\begin{widetext}
\begin{equation}
H_{\mathrm{eff}}(k_z) = \\
\begin{pmatrix}
\epsilon_{p, \mathrm{odd}} & 0 & 0 & t_{\perp} \cos(\frac{k_zc}{2}) & it_{\perp} \sin(\frac{k_zc}{2}) & 0 \\
0 & \epsilon_{p, \mathrm{even}} & t_{\parallel} & -it_{\perp} \sin(\frac{k_zc}{2}) & -t_{\perp} \cos(\frac{k_zc}{2}) & 0 \\
0 & t_{\parallel} & \epsilon_{d} & 0 & 0 & 0 \\
t_{\perp} \cos(\frac{k_zc}{2}) & it_{\perp} \sin(\frac{k_zc}{2}) & 0 & \epsilon_{p, \mathrm{odd}} & 0 & 0 \\
-it_{\perp} \sin(\frac{k_zc}{2}) & -t_{\perp} \cos(\frac{k_zc}{2}) & 0 & 0 & \epsilon_{p, \mathrm{even}} & t_{\parallel} \\
0 & 0 & 0 & 0 & t_{\parallel} & \epsilon_{d}
\end{pmatrix}.
\label{EqH}
\end{equation}
\end{widetext}

Fig. \ref{Fig4}(a) (right half) shows the six bands computed from Eq. \ref{EqH}. When $k_z$ = 0, the eigenvalues can be solved analytically (see Appendix B) and the orbital compositions of the eigenstates are simple [Fig. \ref{Fig4}(b)]. The S $3p_{z, \mathrm{odd}}$ orbitals form bonding and antibonding combinations with respect to the vdW gap, $\sigma_{\textrm{odd}}$, $\sigma^*_{\textrm{odd}}$, with average energy $\epsilon_{p, \mathrm{odd}}$ and bandwidth $w_{\textrm{odd}} = 2t_{\perp}$. Comparison with experiment yields $t_{\perp}$ = 1.34 eV and $\epsilon_{p, \mathrm{odd}}$ = $-$1.75 eV (Fig. \ref{Fig4}(a); left half). The even orbitals S $3p_{z, \mathrm{even}}$ and Nb $4d_{z^2}$ hybridize and produce two pairs of bonding and antibonding states, $\sigma_{\textrm{even1}}$, $\sigma^*_{\textrm{even1}}$ and $\sigma_{\textrm{even2}}$, $\sigma^*_{\textrm{even2}}$. If there were no hybridization, the bandwidths of the independent pairs of $p_{z, \mathrm{even}}$ and $d_{z^2}$ bands would have been $2t_{\perp}$ and zero, respectively. With sizeable hybridization, the bandwidths are given approximately by
\begin{equation}
w_{\textrm{even1}}, w_{\textrm{even2}} \approx t_{\perp}[1\mp (\epsilon_{p, \mathrm{even}} - \epsilon_{d})/(2t_{\parallel})].
\label{Eqw}
\end{equation}
From experiment, we extract $w_{\textrm{even1}}$ = 1.09 eV, whereas $w_{\textrm{even2}}$ is inaccessible because $\sigma_{\textrm{even2}}$, $\sigma^*_{\textrm{even2}}$ lie above $E_F$. 

\subsection{Comparison with DFT} 

\begin{figure*}
\includegraphics[scale=1]{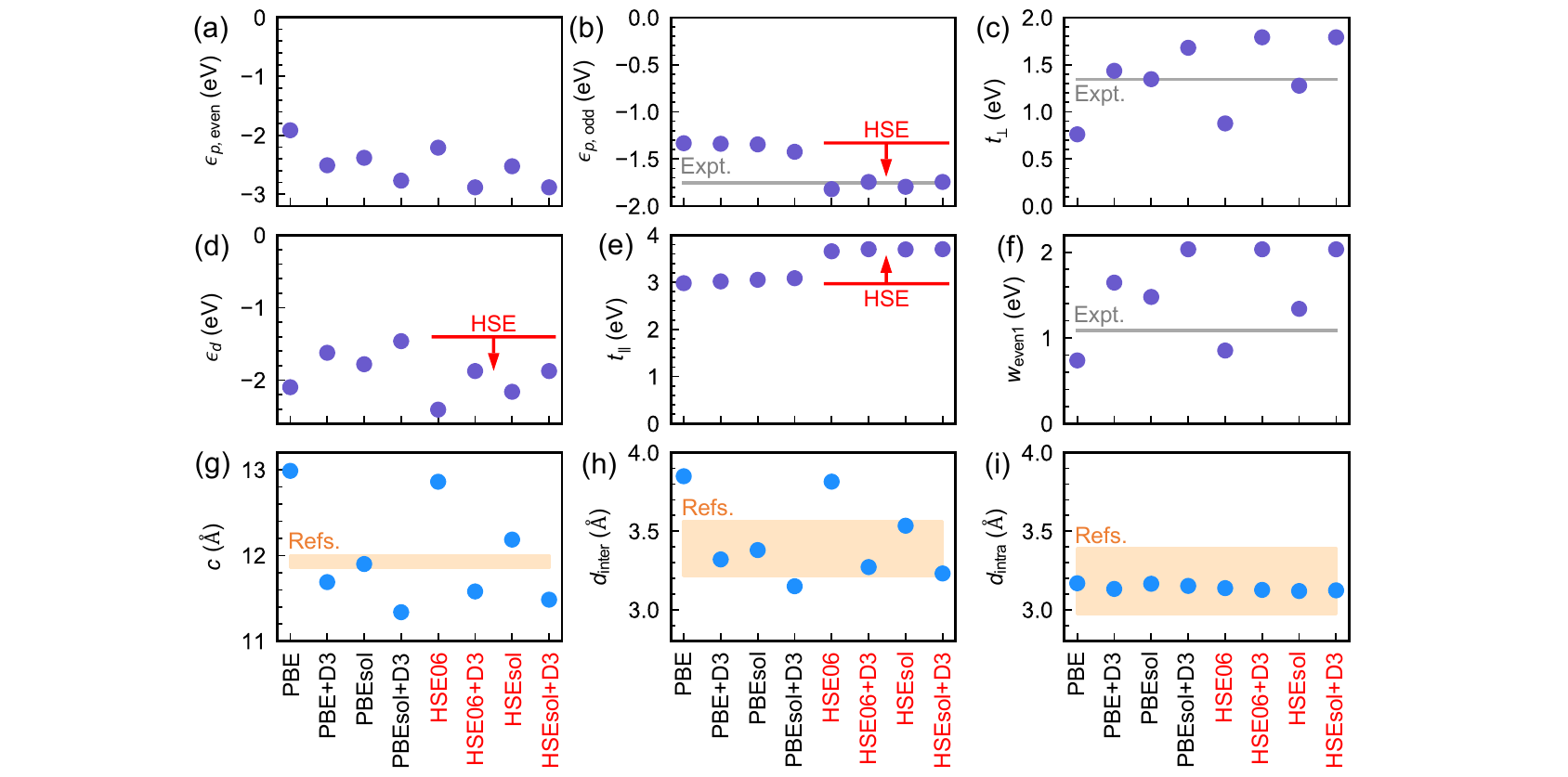}
\caption{Benchmark of DFT methods against interlayer coupling in NbS$_2$. (a)--(f) TB parameters derived from fits of Eq. (\ref{EqH}) to DFT calculations with various exchange-correlation and hybrid functionals. Experimental values are depicted as gray horizontal lines. The red horizontal lines and arrows denote a systematic offset introduced by hybrid functionals. (g)--(i) Optimized structural parameters corresponding to the DFT calculations. The shaded orange regions represent the spread of structural parameters reported in Refs. \cite{Jellinek_Nature_1960, Fisher_IC_1980, Pfalzgraf_JPF_1987, Carmalt_JMC_2004}. The intralayer S-S distance ($d_{\textrm{intra}}$) and the interlayer S-S distance ($d_{\textrm{inter}}$) are defined in Fig. \ref{Fig1}(a).}
\label{Fig5}
\end{figure*}

Since only four of the six bands lie below $E_F$, a full fit of the experimental data to Eq. (\ref{EqH}) with five parameters is underconstrained, and we cannot uniquely determine the experimental values of $\epsilon_{p, \mathrm{even}}$, $\epsilon_{d}$, and $t_{\parallel}$. To estimate these parameters, we fit Eq. (\ref{EqH}) to DFT calculations, an example of which is shown in Fig. \ref{Fig4}(a) (right half). Given previous reports emphasizing the use of vdW, $GW$, and other many-body corrections in modeling NbS$_2$~\cite{Heil_PRL_2017, Heil_PRB_2018, vanLoon_npj_2018, Wang_CPB_2020, ElYoubi_PRB_2021}, we explore a range of functionals in Figs. \ref{Fig5}(a)--\ref{Fig5}(f): the standard Perdew-Burke-Enzerhof (PBE) parametrization \cite{Perdew_PRL_1996} of the generalized gradient approximation, the Heyd-Scuseria-Ernzerhof (HSE06) short-range separated hybrid functional \cite{Krukau_JCP_2006}, and the corresponding functionals revised for solids (PBEsol \cite{Perdew_PRL_2008}, HSEsol \cite{Schimka_JCP_2011}). VdW interactions were taken into account by adding dispersion terms according to the D3 method \cite{Grimme_JCP_2010} to the DFT energies (PBE+D3, HSE06+D3, PBEsol+D3, HSEsol+D3). For each calculation, we perform full structural relaxation, since literature values exhibit some ambiguity (Refs. \cite{Jellinek_Nature_1960, Fisher_IC_1980, Pfalzgraf_JPF_1987, Carmalt_JMC_2004}; see Supplementary Table 4). Values for the $c$ lattice constant and intralayer and interlayer S-S distances ($d_{\textrm{intra}}$, $d_{\textrm{inter}}$) are plotted in Figs. \ref{Fig5}(g)--\ref{Fig5}(i). 

In Fig. \ref{Fig5}, we can understand the variation in the $c$-axis parameter as follows: PBE and HSE06, which are based on the same exchange functional without and with admixture of Hartree-Fock exchange, miss the vdW interactions and underbind the NbS$_2$ layers, leading to larger $c$. PBEsol and HSEsol perform reasonably in estimating $c$. The D3 corrections overbind the NbS$_2$ layers, leading to smaller $c$. Interestingly, $d_{\textrm{intra}}$ remains relatively constant, whereas $d_{\textrm{inter}}$ varies according to $c$. However, it should be noted that the potential energy surface is very flat along $c$, and for a wide range of $d_{\textrm{inter}}$, the total energy varies only by roughly 30 meV~\cite{SM}. 

From Fig. \ref{Fig5}, we can also determine which TB parameters are sensitive to atomic positions, which TB parameters are sensitive to the screening in the hybrid functionals, and which TB parameters are sensitive to both. The interlayer hopping $t_{\perp}$ [Fig. \ref{Fig5}(c)] and bandwidth $w_{\textrm{even1}}$ [Fig. \ref{Fig5}(f)] show the greatest variation across calculations and vary inversely proportional to $c$ [Fig. \ref{Fig5}(g)] and $d_{\textrm{inter}}$ [Fig. \ref{Fig5}(h)], as expected. On the other hand, $\epsilon_{p, \mathrm{odd}}$ and $t_{\parallel}$ are bimodal, adopting one value for PBE functionals, and another for hybrid functionals [red arrows and bars in Figs. \ref{Fig5}(b) and \ref{Fig5}(e)], with little dependence on $c$. The parameter $\epsilon_{p, \mathrm{even}}$ [Fig. \ref{Fig5}(a)] shows moderate variation directly proportional to $c$, whereas $\epsilon_{d}$ [Fig. \ref{Fig5}(d)] is influenced by both $c$ and hybrid functionals. Although we might expect the intralayer terms to be insensitive to interlayer distances, two factors could account for the observed $c$-dependence of $\epsilon_{p, \mathrm{even}}$ and $t_{\parallel}$: First, intralayer hopping is indirectly affected by the $c$ lattice constant through concomittant changes in the $a$ lattice constant. Second, there is an additional interlayer hopping channel between the S 3$p_z$ and Nb 4$d_{z^2}$ orbitals, which is sensitive to $c$. These effects may be buried in $\epsilon_{p, \mathrm{even}}$ and $t_{\parallel}$. 

\section{Discussion} 

The extraction of interlayer hopping parameters in this work hinges upon the SX-ARPES data with $k_z$ dependence. In previous ARPES measurements of NbS$_2$ acquired at a fixed photon energy~\cite{Sirica_PRB_2016, Heil_PRB_2018, ElYoubi_PRB_2021}, the corresponding value of $k_z$ was unknown. Furthermore, due to matrix element effects, the $\sigma_{\textrm{even2}}$, $\sigma^*_{\textrm{even2}}$ bands needed to estimate $t_{\perp}$ are visible at different photon energies. For example, the $\sigma^*_{\textrm{even2}}$ band was not detected at $h\nu$ = 40 eV.

By visual inspection of previously reported SX-ARPES data of TMDC compounds with Se, we can similarly estimate the interlayer hopping $t_{\perp}$ between overlapping Se 4$p_z$ orbitals to be roughly 1.3--1.4 eV in 2$H$-NbSe$_2$ \cite{Weber_PRB_2018} and 1.5 eV in 1$T$-VSe$_2$ \cite{Strocov_PRL_2012}. While the larger size of the Se 4$p_z$ orbitals compared to the S 3$p_z$ orbitals would increase interlayer coupling, a larger $d_{\textrm{inter}}$ would decrease interlayer coupling, but $d_{\textrm{inter}}$ is difficult to determine precisely: 3.22--3.56 \AA~for NbS$_2$ \cite{Jellinek_Nature_1960, Fisher_IC_1980, Pfalzgraf_JPF_1987, Carmalt_JMC_2004}, compared to 3.53--3.55 \AA~for NbSe$_2$ \cite{Brown_AC_1965, Marezio_JSSC_1972} and 3.52--3.56 \AA~for VSe$_2$ \cite{Levy_JPC_1979, Hayashi_JSSC_1978}. In any case, the interlayer hopping in these TMDCs is larger than the value of 0.35--0.48 eV observed in few-layer graphene and graphite \cite{Ohta_PRL_2007}. This does not necessarily imply that graphite is electronically more 2D than the TMDCs. In the former, the same C 2$p_z$-like orbitals that form the $\pi$ bands intersecting $E_F$ also constitute the primary channel of interlayer hopping. In the latter, multiple bands with both transition metal $d$ and chalcogen $p$ characters lie near $E_F$, but there is only significant interlayer hopping through the $p_z$ orbitals. The relative weight of $p_z$ orbitals in the composition of the Femi surface is an additional factor that determines the electronic dimensionality of TMDCs.

Overall, with respect to the $\sigma_{\textrm{odd}}$, $\sigma^*_{\textrm{odd}}$ bands closest to $E_F$, the HSEsol calculation best reproduces the experimental NbS$_2$ data and is shown in Figs. \ref{Fig2}(b), \ref{Fig2}(c), \ref{Fig3}(a), and \ref{Fig4}(a), together with the experimental data. In particular, the HSEsol calculation correctly predicts the band maximum of $\sigma^*_{\textrm{odd}}$ lying 0.4 eV below $E_F$. Previous calculations with PBE, $GW$ corrections, and/or vdW corrections either predict this band crossing $E_F$ and forming an additional hole pocket at $\Gamma$, or lying barely below $E_F$ \cite{Heil_PRL_2017, Heil_PRB_2018, Wang_CPB_2020}. The implications on superconducting properties are as follows: Calculations of $T_c$ using DFT band structures with a $\sigma^*_{\textrm{odd}}$ Fermi pocket overestimate the density of states around $E_F$ and exceed the experimental $T_c$ by a factor of 2-3 \cite{Heil_PRL_2017}. Our results may yield improved estimates of $T_c$. The absence of S 3$p_z$ states with strong $k_z$ dispersion at $E_F$ also causes the Fermi surface to be more 2D, which may contribute to the observed enhancement of the in-plane upper critical field, as orbital depairing is suppressed. In contrast, the Se 4$p_z$ states form a small 3D pocket in NbSe$_2$~\cite{Weber_PRB_2018}, which may partly explain the reduced anisotropy in its upper critical field. Our results also suggest an avenue to raising $T_c$ in NbS$_2$ by applying uniaxial pressure along the $c$-axis, which enhances the interlayer hopping $t_{\perp}$ and pushes the band maximum of $\sigma^*_{\textrm{odd}}$ closer to $E_F$.

Nevertheless, the HSEsol calculation does not perform as well in reproducing the $\sigma_{\textrm{even1}}$, $\sigma^*_{\textrm{even1}}$ bands and the $k_z$-nondispersive bands [Fig. \ref{Fig3}(a)]. More sophisticated theoretical techniques, such as dynamical mean-field theory, could provide one remedy \cite{Kamil_JPCM_2018}. Alternatively, these discrepancies may arise from the presence of defects in real materials. 2$H$-NbS$_2$ and related compounds are prone to stacking faults, polytypism, Nb intercalants, and S vacancies, which sensitively affect the structural and hence interlayer hopping parameters \cite{Lian_PhysicaC_2017, Leroux_PRB_2018, Wen_PRB_2020}. In our SX-ARPES data, we note some faint flat bands (e.g., around $-$4 eV in Fig. \ref{Fig2}) that may be defect levels \cite{Sirica_PRB_2016}. The RRR value of 11 and XPS measurements of core S 2$p$ levels~\cite{SM} also corroborate the presence of defects -- specifically, Nb intercalants. Thus, direct experimental probes of interlayer coupling are useful in these scenarios.    

\section{Conclusion}

We have utilized SX-ARPES to probe the interlayer coupling of superconducting 2$H$-NbS$_2$. The extraction of interlayer hopping parameters is aided by the use of an effective TB model, which also facilitates the comparison with DFT calculations using various functionals. An interesting extension of our work would be to measure the $k_z$ dispersion in superlattice heterostructures with monolayer NbS$_2$ separated by buffer layers, realized either through misfit compounds \cite{Devarakonda_Science_2020} or thin films. Such structures hold promise for realizing exotic superconductivity in the presence of an in-plane magnetic field, and the interlayer coupling between NbS$_2$ layers is a key parameter.

\begin{acknowledgments}

We thank K. Pflaum for technical support. We acknowledge the Paul Scherrer Institut, Villigen, Switzerland, for provision of synchrotron radiation beam time at the ADRESS beamline of the Swiss Light Source. D.H. acknowledges support from a Humboldt Research Fellowship for Postdoctoral Researchers. N.B.M.S. acknowledges partial financial support from Microsoft.

\end{acknowledgments}

\section*{Appendix A: Methods}

Superconducting single crystals of 2$H$-NbS$_2$ were purchased from 2D Semiconductors. Resistivity measurements in the standard four-probe configuration were performed in a physical property measurement system (PPMS) with a 14 T magnet and rotation stage (Quantum Design). 

SX-ARPES measurements were performed at the ADRESS beamline of the Swiss Light Source. The samples were cooled down to 12--14 K in a transfer chamber with 10$^{-10}$ mbar pressure, cleaved to expose a fresh surface, then transferred to an analysis chamber with 10$^{-11}$ mbar pressure. The photon energy range used for Fig. \ref{Fig3} was 350--700 eV. The pass energy of the PHOIBOS-150 analyzer was set to 80 eV, and the corresponding angle-resolving mode was medium angle mode ($\pm$9$^{\circ}$). 

The experiment geometry is reported in the Supplemental Material of Ref. \cite{Strocov_PRL_2012}, along with the corresponding formulas used to determine the electron momentum. Corrections due to the incident photon momentum were taken into account. The conversion from photon energy $h\nu$ to momentum $k_z$ was determined by assuming a free-electron-like final state, with inner potential chosen to match the experimentally observed band periodicity. These calculations were performed using the MATools code.

In addition to momentum-resolved measurements of the valence bands, we also performed XPS measurements of the core levels in NbS$_2$~\cite{SM}. Since the spectrum of a given core level exhibited shifts when taken at different photon energies due to an error in photon energy calibration, we made sure to correct these shifts by aligning to the Fermi edge of a valence band spectrum taken at the identical photon energy. 

XPS spectra were processed and analyzed using the CASAXPS software. For the S 2$p$ core level spectrum, we fitted the peaks to Gaussian-Lorentzian mixture functions on top of a Shirley background. We constrained the doublet spacing to the literature value of 1.18 eV \cite{Moulder_1992} and the area ratio of the doublets to 1:2 ($p$ orbitals). For the Nb 3$d$ core level spectrum, we constrained the doublet spacing to 2.8 eV, which is within the range of values (2.7--2.9 eV \cite{Wang_NatComm_2017, Lin_NatMat_2019}) reported for NbSe$_2$, and the area ratio of the doublets to 2:3 ($d$ orbitals). Due to the asymmetry in one of the peaks, we employed a Doniach-Sunjic line shape \cite{Morris_PRB_2000}. Photoionization cross sections were interpolated from theoretical computed values by Yeh and Lindau \cite{Yeh_ADNDT_1985}.

DFT calculations were performed using the CRYSTAL17 code~\cite{Dovesi_WIRE_2018}. The scalar relativistic pseudopotentials and the Gaussian-type basis functions are given in the Supplemental Material \cite{SM} (see, also, Refs.~\cite{Andrae_TCA_1990, Bergner_MP_1993, Martin_JCP_2001} therein). Spin-orbit coupling was not included, since the spin splitting is small compared to the experimental SX-ARPES resolution. The mesh of $k$-points was determined by applying shrinking factors of 12 along $a^*$ and $b^*$ and 4 along $c^*$. This gives rise to 57 $k$-points in the irreducible part of the Brillouin zone. Starting from the experimental structure~\cite{Jellinek_Nature_1960}, all free parameters within the given space group, the lattice parameters $a$ and $c$, as well as $z$(S), were relaxed until the root mean square on the gradient was below 0.0001 a.u. and the root mean square on the displacement was below 0.0004 a.u.

Atomic structures were visualized using VESTA \cite{Momma_JAC_2011}.

\section*{Appendix B: TB Model}

At $k_z$ = 0, the eigenvalues of Eq. (\ref{EqH}) can be solved analytically:
\begin{equation}
\epsilon_{\sigma_{\mathrm{odd}}, \sigma^*_{\mathrm{odd}}}(k_z = 0) = \epsilon_{p, \mathrm{odd}} \pm t_{\perp}, 
\end{equation}
\begin{equation}
\begin{split}
\epsilon_{\sigma_{\mathrm{even1}}, \sigma^*_{\mathrm{even1}}}(k_z = 0) = \frac{1}{2} \bigg[ \epsilon_{p, \mathrm{even}} + \epsilon_{d} \pm t_{\perp} - \\ \sqrt{(\epsilon_{p, \mathrm{even}} - \epsilon_{d} \pm t_{\perp})^2 + 4 t_{\parallel}^2} \bigg],
\label{Eqe2}
\end{split}
\end{equation}
and
\begin{equation}
\begin{split}
\epsilon_{\sigma_{\mathrm{even2}}, \sigma^*_{\mathrm{even2}}}(k_z = 0) = \frac{1}{2} \bigg[ \epsilon_{p, \mathrm{even}} + \epsilon_{d} \pm t_{\perp} + \\ \sqrt{(\epsilon_{p, \mathrm{even}} - \epsilon_{d} \pm t_{\perp})^2 + 4 t_{\parallel}^2} \bigg].
\label{Eqe3}
\end{split}
\end{equation}
From Eqs. (\ref{Eqe2}) and (\ref{Eqe3}), we can extract simple expressions for the bandwidths $w_{\textrm{even1}}$, $w_{\textrm{even2}}$ in a few limiting cases. When there is no hybridization, i.e., $t_{\parallel}$ = 0, the bandwidth of the S $3p_{z, \mathrm{even}}$ states is $2t_{\perp}$ and that of the Nb $4d_{z^2}$ states is zero. When $\epsilon_{p, \mathrm{even}}$ = $\epsilon_{d}$ and $t_{\parallel}$ is nonzero, $w_{\textrm{even1}}$ = $w_{\textrm{even2}}$ = $t_{\perp}$. When $(\epsilon_{p, \mathrm{even}} - \epsilon_{d} \pm t_{\perp} )^2 / 4 t_{\parallel}^2 \ll 1$, we can apply a truncated binomial expansion to the square roots in Eqs. (\ref{Eqe2}) and (\ref{Eqe3}) and obtain
\begin{equation}
 w_{\textrm{even1}}, w_{\textrm{even2}} \approx t_{\perp}\bigg[ 1 \mp \frac{\epsilon_{p, \mathrm{even}} - \epsilon_{d}}{2t_{\parallel}} \bigg].
\end{equation}
From density functional theory (DFT) calculations, we have confirmed that $(\epsilon_{p, \mathrm{even}} - \epsilon_{d} \pm t_{\perp} )^2 / 4 t_{\parallel}^2 $ ranges from 0.004 to 0.23 and hence, the binomial approximation is reasonable here.


%

\clearpage

\onecolumngrid

\setcounter{figure}{0}
\setcounter{equation}{0}
\setcounter{table}{0}
\setcounter{section}{0}
\setcounter{subsection}{0}
\makeatletter
\renewcommand{\thefigure}{S\@arabic\c@figure}
\renewcommand{\theequation}{S\@arabic\c@equation}
\renewcommand{\thetable}{S\@arabic\c@table}
\newcounter{SIfig}
\renewcommand{\theSIfig}{S\arabic{SIfig}}
\newcounter{SIeq}
\renewcommand{\theSIeq}{S\arabic{SIeq}}

\section*{Supplemental Material}

\subsection*{Supplementary Note 1: X-ray Photoelectron Spectroscopy (XPS)}

\begin{figure}[h]
\includegraphics[scale=1]{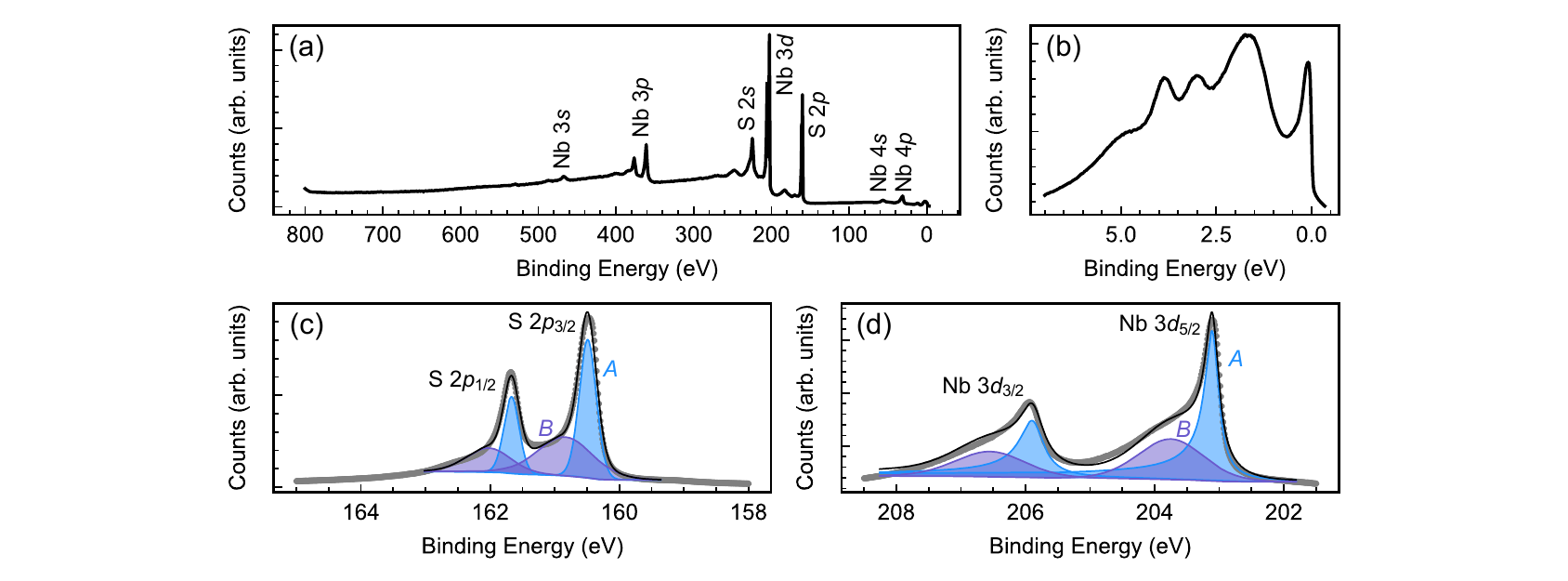}
\caption{XPS measurements of NbS$_2$. (a) Survey spectrum with prominent core levels labeled. (b) Valence band spectrum. (c) S $2p_{1/2}$ and $2p_{3/2}$ doublet. (d) Nb $3d_{3/2}$ and $3d_{5/2}$ doublet. In (c) and (d), the gray circles are the measured data, the black line is the overall fit, and the shaded areas are the individual peaks that constitute the fit. Photon energies: 1000 eV for (a), 402.5 eV for (b)--(d).}
\refstepcounter{SIfig}
\label{FigS1}
\end{figure}

Figure \ref{FigS1}(a) shows the survey spectrum with the expected core levels for Nb and S. Figure \ref{FigS1}(b) shows the momentum-integrated valence band spectrum, which is qualitatively similar to that in Ref. \cite{Sirica_PRB_2016}. 

Figure \ref{FigS1}(c) shows the spectrum of the S 2$p_{1/2}$ and 2$p_{3/2}$ doublet. Fit parameters are provided in Table \ref{T_XPS}. We observe that each doublet consists of two components, labeled $A$ and $B$, indicating two species of S. This observation is reminiscent of XPS measurements in Fe$_x$NbS$_2$ \cite{Saitoh_JES_2005}, where the presence of the Fe intercalant creates two crystallographically inequivalent S sites. In those reports, the peak with lower binding energy ($A$ in our case) was ascribed to S atoms in pristine sites of 2$H$-NbS$_2$, whereas the peak with higher binding energy ($B$ in our case) was ascribed to S atoms coordinated with intercalants. The area ratio of the peaks was roughly consistent with the expected intercalant concentration $x$. For example, at $x$ = 1/3, every S atom is coordinated with an intercalant, and hence the peak with lower binding energy was mostly suppressed.    

\begin{table}[h]
\center
\setlength{\tabcolsep}{7pt}
\caption{XPS fit parameters for the S $2p$ and Nb $3d$ core levels. GL($p$) stands for a Gaussian-Lorentzian mixture function, where $p$ is the mixing factor. $p$ = 100 represents a pure Lorentzian, whereas $p$ = 0 represents a pure Gaussian. DS($a$) stands for the asymmetric Doniach-Sunjic line shape, where $a$ is the asymmetry parameter. The area of each peak is reported as a percentage of the total area. FWHM stands for the full width at half maximum.}
\begin{tabular}{c|cc|cc|cc|cc}
\hline \hline
& \multicolumn{2}{c|}{\underline{S $2p_{3/2}$}} &  \multicolumn{2}{c|}{\underline{S $2p_{1/2}$}} & \multicolumn{2}{c|}{\underline{Nb $3d_{5/2}$}} & \multicolumn{2}{c}{\underline{Nb $3d_{3/2}$}} \\
& S$^A$ & S$^B$ & S$^A$ & S$^B$ & Nb$^A$ & Nb$^B$ & Nb$^A$ & Nb$^B$ \\
\hline
Line shape & GL(30) & GL(30) & GL(30) & GL(30) & DS(0.15) & GL(30) & DS(0.15) & GL(30) \\
Area [\%] & 34.75 & 31.91 & 17.38 & 15.96 & 42.85 & 17.03 & 28.71 & 11.41 \\
FWHM [eV] & 0.31 & 1.02 & 0.28 & 0.82 & 0.24 & 1.19 & 0.43 & 1.27\\
Position [eV] & 160.49 & 160.84 & 161.67 & 162.02 & 203.13 & 203.75 & 205.93 & 206.55\\
\hline \hline
\end{tabular}
\label{T_XPS}
\end{table}

We thus may infer the presence of Nb intercalants in our measured NbS$_2$ crystals. As discussed in Ref. \cite{Lian_PhysicaC_2017}, during sample growth or annealing, Nb atoms that lose S may intercalate between layers. From the fits, we find that peak $A$ ($B$) occupies 52\% (48\%) of the total area of the S 2$p$ core spectra. Since every intercalant is coordinated with six neighboring S atoms, and each unit cell of $2H$-NbS$_2$ consists of two NbS$_2$ layers, the concentration of Nb intercalants is one third of the fractional area occupied by peak $B$. Thus, we infer that our sample is Nb$_{1+x}$S$_2$, with a rough estimate of $x \approx 0.48 / 3 = 0.16$. 

Figure \ref{FigS1}(d) shows the spectrum of the Nb 3$d_{3/2}$ and 3$d_{5/2}$ doublet. Again, each doublet is composed of two components, labeled $A$ and $B$, but $A$ is clearly asymmetric, showing a long tail towards higher binding energies. Two possible interpretations exist for peaks $A$ and $B$. They could represent two species of Nb, one within the NbS$_2$ layer, and the other as an intercalant. However, in both cases, the Nb atom is coordinated with six S atoms, so their local environment may be rather similar. Alternatively, there is some discussion of similar observations in Fe$_x$NbS$_2$ \cite{Saitoh_JES_2005}, Cr$_{1/3}$NbS$_2$ \cite{Sirica_PRB_2016}, and Nb-doped TiO$_2$ \cite{Morris_PRB_2000}, where the observation of two components per Nb doublet does not indicate two Nb species, but a final-state screening effect. Essentially, the core hole produced during photoexcitation may pull down some localized $d$ states below $E_F$, leading to a ``well-screened'' final state with asymmetric lineshape ($A$), in addition to the ``poorly screened'' final state with symmetric lineshape ($B$). To fit the asymmetric $A$ peak, we employed a Doniach-Sunjic line shape \cite{Morris_PRB_2000}.

\begin{figure}[h]
\includegraphics[scale=1]{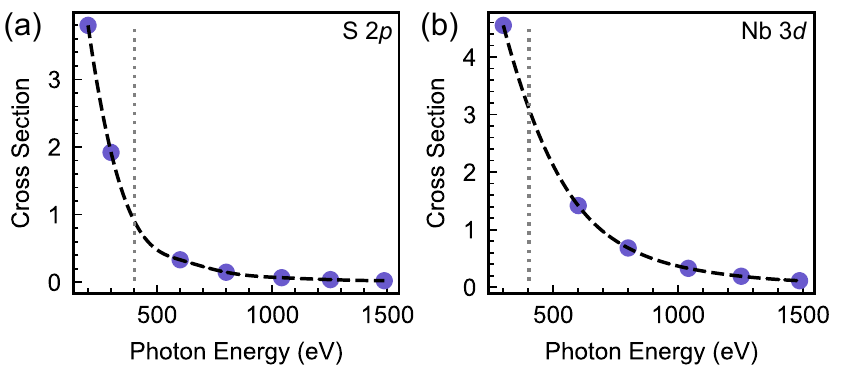}
\caption{XPS photoionization cross sections. The circles represent computed values by Yeh and Lindau~\cite{Yeh_ADNDT_1985} for (a) S 2$p$ and (b) Nb 3$d$ core levels. The black dashed lines represent cubic spline interpolations necessary to estimate the cross section at the photon energy used, 402.5 eV (gray dotted lines).}
\label{FigS2}
\end{figure}

The relative area ratio of the S 2$p$ and Nb 3$d$ core levels gives us another rough estimate of the stoichiometry. We computed the total area of the S 2$p$ and Nb 3$d$ peaks, then normalized by the theoretical photoionization cross sections computed by Yeh and Lindau \cite{Yeh_ADNDT_1985}. The cross sections are dependent on photon energy, so for 402.5 eV, we performed cubic spline interpolation of the available tabulated data (Fig.~\ref{FigS2}) and estimated the cross sections to be 0.89 for S 2$p$ and 3.1 for Nb 3$d$. This yields an $x$ value of 0.14 for Nb$_{1+x}$S$_2$, which is close to our first estimate.

\subsection*{Supplementary Note 2: Photon Polarization Dependence}

\begin{figure}[h]
\includegraphics[scale=1]{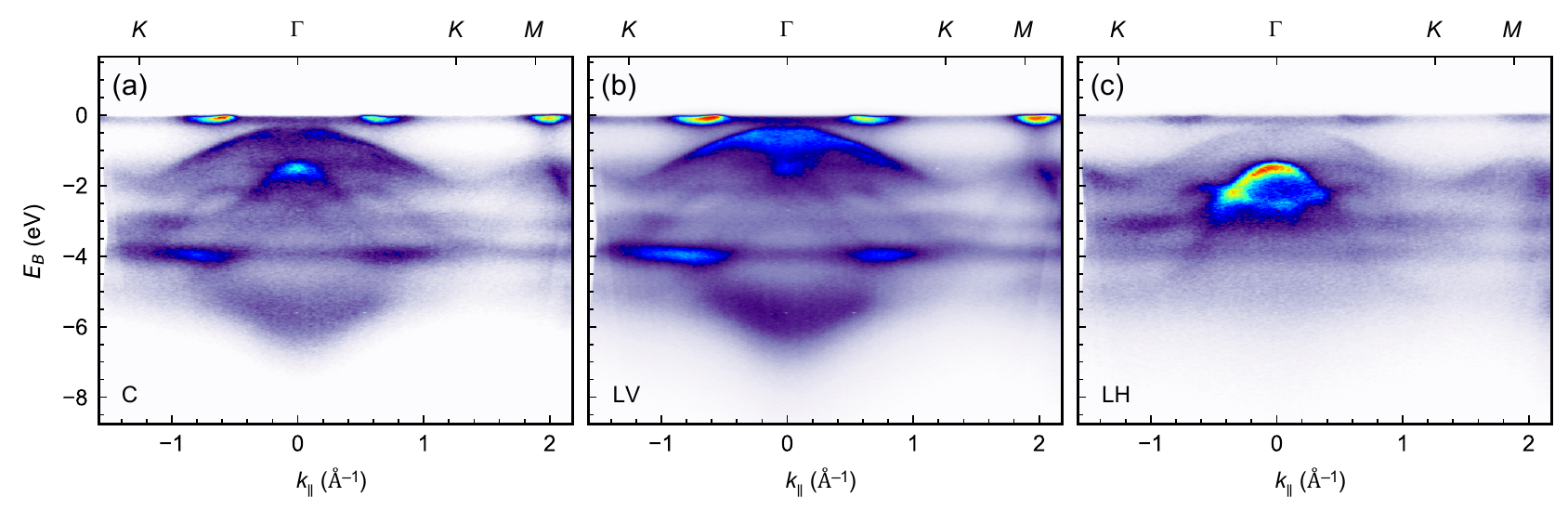}
\caption{Photon polarization dependence. SX-ARPES intensity cuts in the $k_z$ = 0 plane using (a) circularly, (b) linearly vertically, and (c) linearly horizontally polarized photons. $h\nu$ = 590 eV.}
\label{FigS3}
\end{figure}

Figure \ref{FigS3} shows soft-x-ray angle-resolved photoemission spectroscopy (SX-ARPES) intensity cuts along $\Gamma$--$K$--$M$ taken with circularly and linearly polarized photons. The different contrasts of the bands reflect their different orbital characters and are very similar to measurements on NbSe$_2$~\cite{Weber_PRB_2018}. Photons with linear vertical (LV) polarization, i.e., $p$ polarization, pick out contributions from orbitals that are even with respect to the $k_x$-$k_z$ plane, or with lobes extending out of the NbS$_2$ layer \cite{Weber_PRB_2018, Ugeda_NatPhys_2015}. Bands with enhanced intensity using LV photons include the $k_z$-dispersive bands at $\Gamma$ (around $-$0.5 and $-$6 eV) and the near-$E_F$ bands along $\Gamma$--$K$ and at $M$, all of which contain dominant contributions from Nb 4$d_{z^2}$ and/or S 3$p_z$ orbitals [Fig. \ref{FigS3}(b)]. Photons with linear horizontal (LH) polarization, i.e., $s$ polarization, pick out contributions from orbitals that are odd with respect to the $k_x$-$k_z$ plane, or with lobes confined within the NbS$_2$ layer. Bands with enhanced intensity using LH photons include the $k_z$-nondispersive bands at $\Gamma$ (around $-$1.5 and $-$2.5 eV), which contain dominant contributions from Nb 4$d_{xz/yz}$, Nb 4$d_{xy/x^2-y^2}$, and/or S 3$p_{x/y}$ orbitals [Fig. \ref{FigS3}(c)]. All bands are reasonably visible with circularly polarized (C) photons. Unless otherwise indicated, all reported SX-ARPES data were acquired with C polarization.

\subsection*{Supplementary Note 3: DFT}

Tables \ref{Tpseudo} and \ref{Tcomp} present computational details and parameters from the CRYSTAL17 code.

\begin{table}[h]
\setlength{\tabcolsep}{6pt}
\caption{Pseudopotentials (scalar relativistic) and basis sets.}
\begin{tabular}{c|cc|cc}
\hline \hline
& \multicolumn{2}{c|}{Nb} & \multicolumn{2}{c}{S} \\
\hline
Pseudopotential & & & & \\
Core & \multicolumn{2}{c|}{[Ar]3$d^{10}$} & \multicolumn{2}{c}{[Ne]} \\
Valence & \multicolumn{2}{c|}{$4s^24p^64d^35s^2$} & \multicolumn{2}{c}{$3s^23p^4$} \\
Ref. & \multicolumn{2}{c|}{\cite{Andrae_TCA_1990}} & \multicolumn{2}{c}{\cite{Bergner_MP_1993}} \\
\hline
& Exponent & Coefficient & Exponent & Coefficent \\
\hline
$s$-shell & 6.566301 & $-$0.8582654 & 6.833518 & $-$0.043875 \\
& 4.586438 & 1.3041672 & 2.077738  & 0.319894 \\
& 3.753770 & 0.5069043 & 0.419121  & $-$0.661233 \\
& & & & \\
& 0.889871 & 1.0 & 0.153237 & 1.0 \\
& & & & \\
& 0.407138 & 1.0 & & \\
& & & & \\
& 0.094271 & 1.0 & & \\
\hline
$p$-shell & 3.070063 & $-$3.9044315 & 1.817139 & $-$0.079227 \\
& 2.237964  & 4.0688070 & 0.855070  & 0.263671 \\
& & & 0.312053 & 0.580682 \\
& 0.852255 & 0.6713910 & & \\
& 0.504436 & 0.3474365 & 0.101687 & 1.0 \\
& & & & \\
& 0.2668 & 1.0 & & \\
& & & & \\
& 0.09 & 1.0 & & \\
\hline
$d$-shell & 4.053563 & $-$0.0204201 & 0.2628 \cite{d-shell} & 1.0 \\
& 1.652600 & 0.2089854 & & \\
& 0.706859 & 0.4705515 & 0.11 & 1.0 \\
& 0.286367 & 0.4758860 & & \\
& & & & \\
& 0.108757 & 1.0 & & \\
\hline
$f$-shell & 0.97 \cite{Martin_JCP_2001} & 1.0 & & \\
& & & & \\
& 0.261 & 1.0 & & \\
\hline \hline
\end{tabular}
\label{Tpseudo}
\end{table}

\begin{table}[h]
\setlength{\tabcolsep}{6pt}
\caption{Tolerance and other computational parameters used in the CRYSTAL17 input.}
\begin{tabular}{ll}
\hline \hline
TOLINTEG & 12 12 12 12 24 \\
TOLPSEUD & 12 \\
TOLDEE & 8 \\
BIPOLAR & 64 64 \\
LEVSHIFT & 6 0 \\
\hline \hline
\end{tabular}
\label{Tcomp}
\end{table}

Structural parameters derived from full ionic relaxation are presented in Table \ref{Tstruc}, along with experimental values from literature.

\begin{table}[h]
\setlength{\tabcolsep}{4pt}
\caption{Structural parameters from DFT calculations and experiments.}
\begin{tabular}{c|cccccccc|cccc}
\hline \hline
& PBE & PBE & PBEsol & PBEsol & HSE06 & HSE06 & HSEsol & HSEsol & Expt. 1 & Expt. 2 & Expt. 3 & Expt. 4 \\
& & +D3 & & +D3 & & +D3 & & +D3 & \cite{Jellinek_Nature_1960} & \cite{Fisher_IC_1980} & \cite{Pfalzgraf_JPF_1987} & \cite{Carmalt_JMC_2004} \\
\hline
$a$ (\AA) & 3.362 & 3.319 & 3.318 & 3.280 & 3.340 & 3.291 & 3.312 & 3.279 & 3.31 & 3.324 & 3.33 & 3.418 \\
$c$ (\AA) & 12.988 & 11.690 & 11.901 & 11.34 & 12.861 & 11.581 & 12.186 & 11.485 & 11.89 & 11.95 & 12.00 & 11.860 \\
$z$(S)/$c$ & 0.128 & 0.116 & 0.117 & 0.111 & 0.128 & 0.115 & 0.122 & 0.114 & 0.125 & 0.108 & 0.125 & 0.1079 \\
\hline \hline
\end{tabular}
\label{Tstruc}
\end{table}

\begin{figure}[h]
\includegraphics[scale=1]{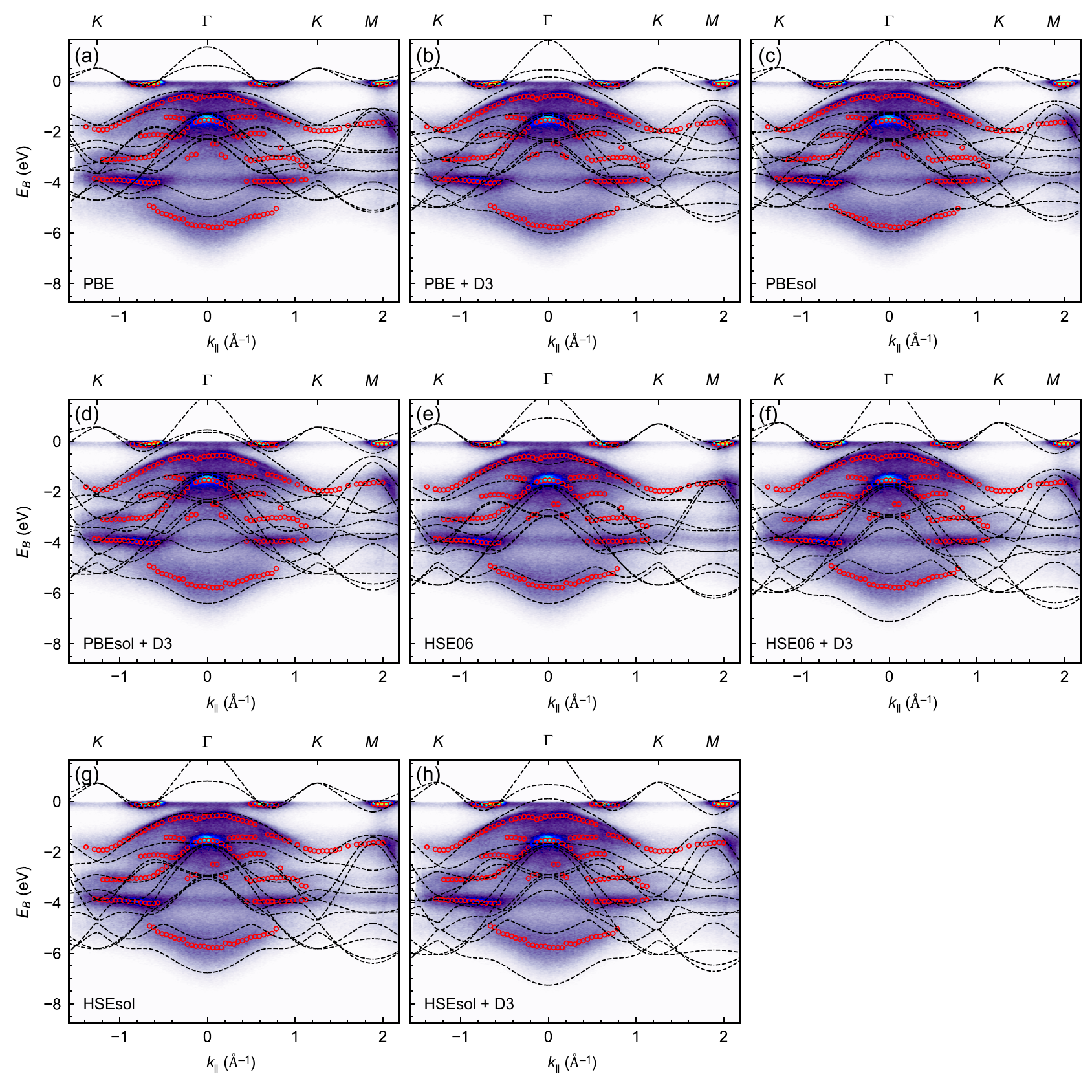}
\caption{Comparison of DFT calculations using different exchange-correlation and hybrid functionals along $\Gamma$--$K$--$M$. The open circles overlaying the experimental SX-ARPES intensities represent local peaks detected in the energy distribution curves, whereas the dashed lines represent the different DFT calculations. $h\nu$ = 590 eV.}
\label{FigS4}
\end{figure}

\begin{figure}[h]
\includegraphics[scale=1]{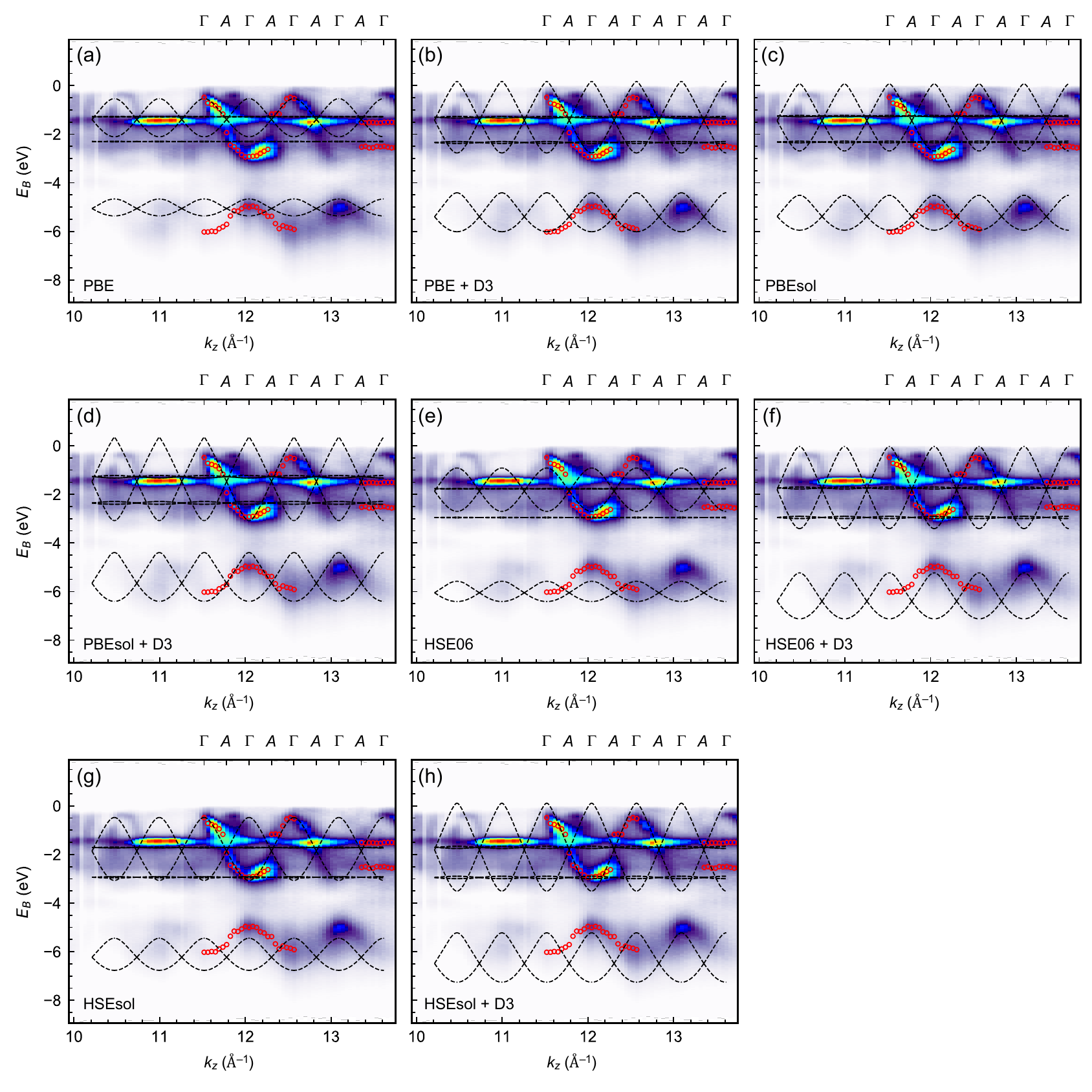}
\caption{Comparison of DFT calculations using different exchange-correlation and hybrid functionals along $\Gamma$--$A$. The open circles overlaying the experimental SX-ARPES intensities represent local peaks detected in the energy distribution curves, whereas the dashed lines represent the different DFT calculations.}
\label{FigS5}
\end{figure}

Figures \ref{FigS4} and \ref{FigS5} show the various DFT calculations overlaying the SX-ARPES intensity along $\Gamma$--$K$--$M$ [reproduced from Fig. 2(b)] and along $\Gamma$--$A$ [reproduced from Fig. 3(a)]. Peaks in the energy distribution curves corresponding to the experimental bands are also overlaid to facilitate the comparison between experiment and theory. As discussed in the main text, the HSEsol calculation performs the best in reproducing the $\sigma_{\textrm{odd}}$, $\sigma^*_{\textrm{odd}}$ bands near $E_F$. However, HSEsol appears to overshoot in lowering the energy of the $\sigma_{\textrm{even1}}$, $\sigma^*_{\textrm{even1}}$ bands, and PBEsol gives a closer match to experiment [Fig.~\ref{FigS5}(c)]. From Fig.~\ref{FigS4}, we also note some trends at points in the Brillouin zone other than $\Gamma$ and $A$. The PBE-based functionals better reproduce the energy of the highest occupied state at $K$, but predict the highest occupied state at $M$ to lie much closer to $E_F$ than experiment. The hybrid functionals, HSE06 and HSEsol in particular, yield better predictions for the highest occupied state at $M$.   

\subsection*{Supplementary Note 4: Dependence on Structural Parameters}

\begin{figure}[h]
\includegraphics[scale=1]{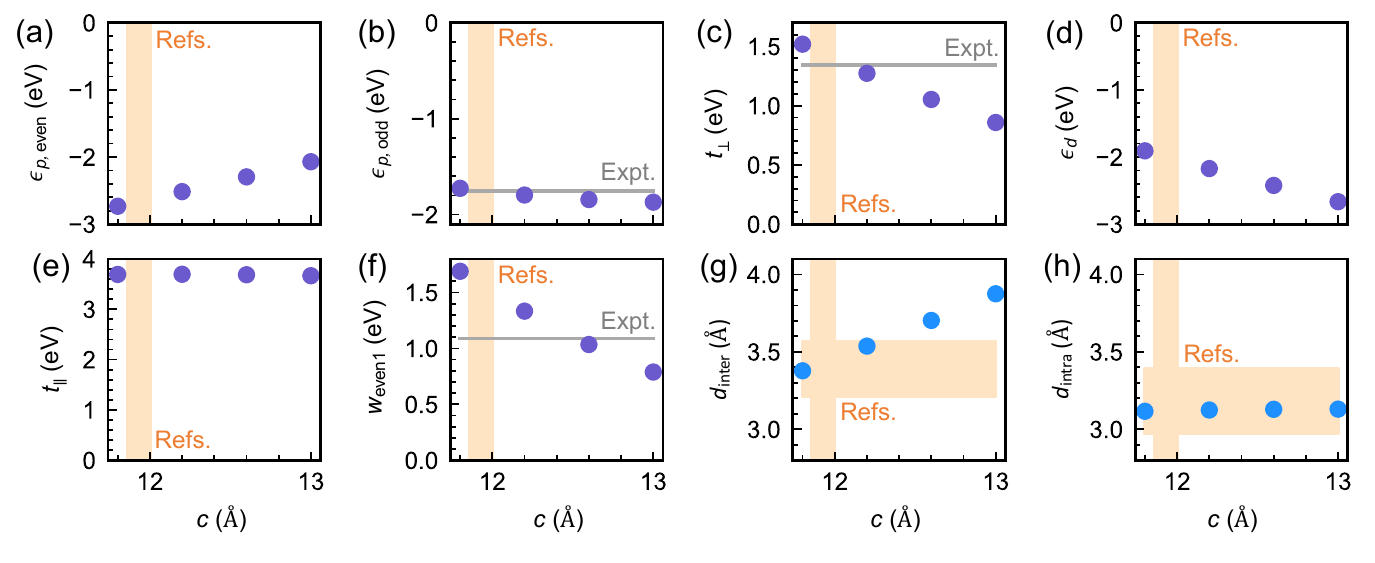}
\caption{Evolution of TB and structural parameters as a function of the $c$ lattice constant. The DFT band structures with the HSEsol functional are computed for $c$ = 11.8, 12.2, 12.6, and 13.0 \AA, then fitted to the TB model. Experimental values extracted from SX-ARPES are depicted as gray horizontal lines. The spread of structural parameters reported in literature \cite{Jellinek_Nature_1960, Fisher_IC_1980, Pfalzgraf_JPF_1987, Carmalt_JMC_2004} is marked by the shaded orange regions.}
\label{FigS6}
\end{figure}

To further elucidate the dependence of the electronic structure of NbS$_2$ on structural parameters, we computed the band structure with the HSEsol functional for four fixed values of $c$: 11.8, 12.2, 12.6, and 13.0 \AA. The other atomic positions were allowed to relax. The structure with $c$ = 12.2 \AA~has the lowest energy, and the energies relative to this structure are +14 meV, +11 meV, and +34 meV for the $c$ = 11.8, 12.6, and 13.0 \AA~structures, respectively. Fig. \ref{FigS6} shows the evolution of the fitted TB and structural parameters as a function of $c$. Consistent with Fig. 5 of the main text, the interlayer hopping $t_{\perp}$ and bandwidth $w_{\textrm{even1}}$ show the greatest variation with $c$, while the orbital energy $\epsilon_{p, \mathrm{odd}}$ and intralayer hybridization $t_{\parallel}$ of even orbitals show the least variation. In terms of the atomic positions, $d_{\textrm{inter}}$ increases as $c$ increases, but $d_{\textrm{intra}}$ retains a relatively constant value. 

\end{document}